# On some peculiarities of Fe - Ni – Al powder mixture properties fabricated by pressing and shock-wave compaction

F.Kh. Akopov[1], I.V. Chkhartishvil[2], V.M. Gabunia[2], G.I. Mamniashvili[1]*,
A.V. Peikrishvili[3], N.M. Chikhradze[3],

[1] E.Andronikashvili Institute of Physics, 0177, Tbilisi, G.Tamarashvili str.
[2] T.Tavadze Institute of Metallurgy and Materials Science, 0177, Tbilisi, 15, Al.Kazbegi av.
[3] Institute of Mining, 0186, Tbilisi, 7, Mindeli, str.

*Corresponding author.
E-mail address: Mamni@iphac.ge
(G.I.Mamniashvili)

## Abstract

It is presented some technological parameters and physical characteristics of Fe - Ni – Al composite materials, fabricated by pressing with following sintering at a high temperature, and also shock-wave compacted using explosives.

The comparison of materials properties fabricated using different technologies at equal precursor powders parameters (dimensions and shape of particles, a way of their mixing and density of filling) points on the difference of their phase compositions, metallographic and defect structure, porosity and some physical properties.

Investigation of compacted metallic powder mixtures has already taken for a long time a leading place in scientific programs [1-4], and their practical application give evident benefit in different branches of industry [5-7]. Besides the known economical benefits with the help of powder technologies one could create articles which differ by their properties from materials fabricated by alloying. Properties of massive powder articles compacted using either way and consisting of particles, which dimensions could vary from part to hundreds of microns, depend on a number of factors. In the first turn these are properties of powder particles themselves, a choosen compacting method, a pressure developed and its duration, sintering temperature of pressed material or value of induced temperature at shock-wave compaction (SWC) caused by interparticle friction and shock compression at passing of a shock-wave, front structure of interparticles contacts, porosity, defect structure, an influence of oxide coating of particles and interaction with environment.

The described systems have doubtlessly features by which they could be counted as being out from the state of thermodynamic equilibrium.

In the case of compaction of powder mixtures from different substances, simultaneously with dislocation effects and selfdiffusion, caused by influence of outer pressing force and forces of capillary origin, it takes place a mass transfer in the subcontact region accompanied by the diffusion adjustment of different atom concentrations. Physical processes of mass transfer, accompanied by the decrease of excessive surface energy, define the densification kinetics at pressing. Chemical processes, depending on temperature and the composition of gas media where sintering takes place, are accompanied by decomposition and restoration of oxides. The sintering peculiarity is related with a recrystallization growing of favorably located particles sometimes essentially exceeding the original characteristic dimension. It is possible when a local value of pressing density appears to be larger than that of nearby regions and is related with the inhomogeneous distribution of stresses developed at pressing [8].

Shock-wave compaction of metallic powder mixtures using explosive substances (ES) makes it possible to control in wide ranges the value of applied pressure, fabricate materials with sufficiently reliable bonds between particles and a high density. Besides the value of pressure, duration of its influence, propagation of shock-wave front and its shape, as well as temperature value induced by interparticle friction and a shock compression, properties of the obtained article depend on the original dimensions of particles and filling density, volumes of charge and explosive substances, the ratio of explosive substance charge height $h_1$ to the height of compacted material $h_2$ and so on.

Changing the $h_1/h_2$ ratio on the expense of the increase of ES charge height it is possible to reach the greater density of material but it should be remembered that possible in this case the excessive increase of detonation velocity could cause the destruction of materials [9].

The explosive compaction results in the improvement of conditions for the creation of particle contacts with metallic bonds improving mechanical properties and specific electroconductivity of materials in the hole as compared with the ones fabricated by pressing.

The choice of the fabrication technology of powder materials and articles from them besides the economical expediency should by defined by the necessity of fabrication of materials with the preset or close to it physico-mechanical properties. In this work it is considered results of the comparison of some properties of the Fe – Ni – Al composite material fabricated by the method of double pressing consequently at 0.6 GPa and 1 GPa with the following sintering in the hydrogen atmosphere and the same powder mixture compacted by the shock-wave compactment (SWC) using the energy of hexagon explosive substance. This comparison allows one to make more clear accent on the necessary corrections for the technology methods of their fabrication from the point of view of the raised problems.

In table 1 it is presented some technological parameters characterizing the fabrication of the investigated compacted materials.

One of parameters which are necessary for the control of the fabricated composite materials is density. Density of the investigated samples was defined by their weighing on the analytical balance VLP-200. Weighing of hair suspended samples was carried out in air and ethyl alcohol. In Fig.1 it is presented the dependences of weight change $\Delta P$ of samples plunged into alcohol on the measurement time t. Weighing of samples it was made during a time necessary for stabilization of the measured weight value.

**Table 1**

| Investigated materials, wt. % | Initial size of grains in $\mu$m | Compacted method | Maximal pressure, GPa | Compacting time | Treatment temperature T, K |
|---|---|---|---|---|---|
| Fe -65 Ni -23 Al -12 | 20-30 | Double pressing + sintering | 0.6 – 1.0 | 0.5 – 1.0 hour | Sintering at 1500 K, 4-5 hours in hydrogen media |
| | | SWC | 20,0 | $10^{-8}$-$10^{-7}$ sec | Selfheating, proportional the energy absorbed during SWC |

Results, presented in Fig.1, point to the fact that the weight record change is caused by a gradual penetration of alcohol into the surface underlying pores. As it is seen from the figure, firstly, the process of pores filling and consequently the weight change is significantly extended in time. Secondly, the weight change takes place in a higher degree with the pressed sample (curve 1).

As far as the method of sample cutting in both cases was identical, the difference in the structure of pores underlying of the sample surface is caused by the compaction technology. As far as the maximal pressure of pressing of the powder charge was twenty times less then pressure developed at SWC (curve 2) the total volume of the surface underlying pores of the pressed material is evidently greater.

The weighing results made it possible to define the density of investigated substances at condition that the any further significant penetration of alcohol into the surface underlying pores and, consequently, the weight change would not take place (Table 2). The presented density values are those taken without allowing for the possible existence of closed pores inside the volume of the investigated material and therefore they are indeed somewhat underestimated. Independently from a way of its fabrication, the volume of possible pores inside the compacted material could be presented as $V_o = m/\rho - V'$, where m is the mass of a sample, $\rho$ is the density of composite materials fabricated by a corresponding compaction method, and V' is the volume occupied by alloy with a similar weight composition of components.

Accordingly the calculation, the density of alloy with the V' volume should be equal to $\rho' = 6,53$ g/cm$^3$. The relative changes of densities $\Delta\rho/\rho$ of pressed and shock-wave compacted samples are presented in Table 2.

The difference of values of the presented quantities points to the presence inside them of the proportional amount of closed pores. The difference in 1.8 % for the shock-wave compacted sample is satisfactory allowing for the fact that the material was fabricated without its heating immediately before SWC, i.e. without the use of hot explosive compaction method (HEC) [3] which, as a rule, results in the higher density values but at the same time the hardness of the article is reduced [3].

**Table 2.**

| Compaction method | Volume of undersurface pores, $V \cdot 10^3$ см$^3$ | $\Delta\rho/\rho'$ , % $\Delta\rho=\rho' - \rho$ | $\rho$, composite density, g/cm$^3$ | $\rho'$, calculated for alloy g/cm$^3$ |
|---|---|---|---|---|
| Pressing + sintering | 1.32 | 12.4 | 5.72 | 6.53 |
| SWC | 1.07 | 1.8 | 6. 41 | |

The X-ray diffraction spectra were obtained by the DRON – 2 installation using CoK$_\alpha$ - irradiation (Fig.2). On the X-ray diffraction spectrum of the sample, fabricated by pressing with a following sintering (Fig.2a), besides reflections from aluminum, iron and nickel it is observed peaks apparently belong to iron and nickel aluminides. From a relatively weak intensity of peaks and the amount of anticipated phases their exact identification is difficult. The X-ray diffraction spectra of sample fabricated by SWC (Fig.2b) didn't show the creation of new chemical compounds.

With a help of scanning electron microscope (SEM) it was established microstructure and spectra of element composition from different regions on the sample surface (Fig.3) compacted by shock-wave. Results of spectral analysis, presented in table 3, revealed the procental distribution of elements in ten different regions on the surface of sample. It is seen from Fig.3 that the investigated sample is a well compacted powder mixture. A high degree of compaction is apparently received as a result of plastic deformation of particles and halfliquid sintering of easily melted aluminum with grains of iron and nickel. This is possible at the adiabatic local heating of aluminum resulting in the temperature increase, induced by interparticle friction and shock pressing. Accordingly the presented microstructure and the results of spectral analyses of the separated regions of sample it is seen compacted powder particles consisted of pure iron and pure nickel as well as the aluminum ones also being clearly seen.

**Table 3.**

Processing option : All elements analyzed (Normalised)

| Spectrum | C | O | Al | Fe | Ni | Total |
|---|---|---|---|---|---|---|
| Spectrum  1 | | | 100.00 | | | 100.00 |
| Spectrum  2 | 27.11 | 9.84 | 50.50 | | 12.55 | 100.00 |
| Spectrum  3 | 40.68 | 12.56 | 38.01 | | 8.75 | 100.00 |
| Spectrum  4 | | | | 100.00 | | 100.00 |
| Spectrum  5 | | | | 100.00 | | 100.00 |
| Spectrum  6 | | | 23.59 | 45.06 | 31.35 | 100.00 |
| Spectrum  7 | 22.39 | 13.56 | 64.05 | | | 100.00 |
| Spectrum  8 | | | | | 100.00 | 100.00 |
| Spectrum  9 | | 12.99 | 61.08 | | 25.94 | 100.00 |
| Spectrum 10 | | | | 100.00 | | 100.00 |
| | | | | | | |
| Max. | 40.68 | 13.56 | 100.00 | 100.00 | 100.00 | |
| Min. | 22.39 | 9.84 | 23.59 | 45.06 | 8.75 | |

From presented in table 3 results of the spectral analysis of the majority of separated regions simultaneously with the identification of aluminum particles it is revealed the presence of carbon and oxygen atoms. The presence of oxygen and carbon on the surface of aluminum particles in the compacted material could be apparently accounted for by the nonhermiticity of exially symmetric storage ampules for the penetration of the exposion production into the charge and a high chemical activity of aluminum, its high reaction capability. The observation of oxygen and carbon atoms in other spectra (spectra 2 and 3) containing besides aluminum atoms also the nickel ones is explaned by the overlapping of scanning area with particles of both these elements. The presented data correlate with ones by Auger spectroscopy method (Fig. 4). In spite of the fact that the X-ray phase analysis didn't show the creation of new phase, there excistence is not excluded in thin undersurface layers of particles.

In Fig. 5 it is presented the temperature spectra of the real part of complex magnetic susceptibility of the pressed sample (Fig.5a) and the sample fabricated by the SWC method (Fig.5b).

Magnetic properties of Fe – Ni – Al powder mixture are revealed as a set of properties of iron, nickel, aluminum particles and located in the region of contacts atomic mixtures of contacting pairs Fe – Ni, Fe – Al, Ni – Al formed at the reduction of the total surface and, consequently, surface energy of the system.

Redistribution of substance in the contact region taking place as a result of applied capillary (Laplace) pressure, proceeds in different ways depending on processes of mutual diffusion and selfdiffusion, defect structure and phase composition [8]. The all above said phenomena doubtlessly influence the domain wall mobility and, as a consequence, the value and temperature dependence of the real part of complex magnetic susceptibility. The heating of sample in the temperature range 80-450K was realized with a constant velocity ~ 1,2 K/min. The kink on the $\chi'(T)$ dependence at 125 K (Fig.5a) completely coincides with the result of the study of similar material described in work [10].

This peculiarity is revealed as phase transition on the temperature spectrum of real part of the complex magnetic susceptibility $\chi'(T)$, and on the temperature spectrum of elasticity modulus $f^2(T)$ and dissipation of sound $Q^{-1}(T)$, described in the pointed work. In the given case it is quite possible that one speaks about magneto-structural or spin-reorientational magnetic phase transitions causing the change of elastic properties through the magnetostriction. From the Fig.5a it is also seen that the spectrum region from 125 K to 380 K could be classified as the region with the relatively weak temperature dependence of the domain walls susceptibility because the domain walls mobility in it is considerably reduced as result of anyone from the possible reasons causing it, starting from the change of phase composition up to the allowing for of the relaxation processes caused by the different types of relaxators.

Above 380 K it is repeatedly observed an intensive rise of $\chi'$ on temperature. In Fig. 5b the $\chi'(T)$ dependence received for sample fabricated by SWC method also could be conditionally divided on three regions as in the previous figures. At heating from the low temperatures it is observed the intensive rise of $\chi'(T)$ but a kink characteristic for phase transitions at 125 K is not observed what apparently points to the absence of this low-temperature phase, characteristic for samples fabricated by pressing technology with a following sintering. The transition to the region with a relatively weak temperature dependence of the domain walls mobility takes place at considerably higher temperature ~ 280 K. In the temperature interval 280- 380 K the rise of $\chi'(T)$ is changed by a small region which is a characteristic for the relaxation process. The region of the following intensive rise of $\chi'(T)$ dependence is displaced into the region of high temperature.

The results of this work point to the fact that at compacting of Fe – Ni – Al powder mixtures by two different ways: the double pressing with a following sintering and the shock-wave compaction at similar original parameters of the compacted powders, porosity, metallographic and defect structures, phase composition and some physico-mechanical properties essentially differ depending on technology of fabrication, namely:

1. as a result of double pressing (P ~ 0.6 – 1.0 GPa) with the following sintering (at 1500 K during 4.5 hours in the hydrogen substance) density reaches 87 %;

2. the shock-wave compaction of powder mixture (P ~ 20 GPa) significantly reduces material's porosity as compared with the pressing and it reaches the value ~ 98 %;
   the reduction of porosity is connected apparently with the peculiarity of SWC technology facilitating the formation of a dense mixture of powder particles of different sizes into the whole volume of the sample, when the space between large particles is filled by the smaller ones what generally could facilitate the increase of the strength of fabricated materials;

3. the high degree of densification at SWC is possibly caused also by the halfliqued sintering of easily melted aluminum with iron and nickel grains at the local heating (in the adiabatic conditions) in the process of shock-wave spreading;

4. from X-ray diffraction spectra it is seen that as a result of pressing with the following sintering at 1500 K it is possible the creation of the iron and nickel aluminides and because of the relatively weak intensity and amount of supposed phases it is difficult to carry out their exact identification;

5. as a result of SWC process it is conserved the original phase composition;

6. the enlargement of a part of grains is a result of the simultaneous influence of a high pressure developed at SWC and a short time but apparently sufficiently significant pulsed increase of temperature;

7. the difference of temperature spectra of the real part of complex magnetic susceptibility of the pressed sample and the one fabricated by SWC method is caused by the difference of their phase composition and defect structures.

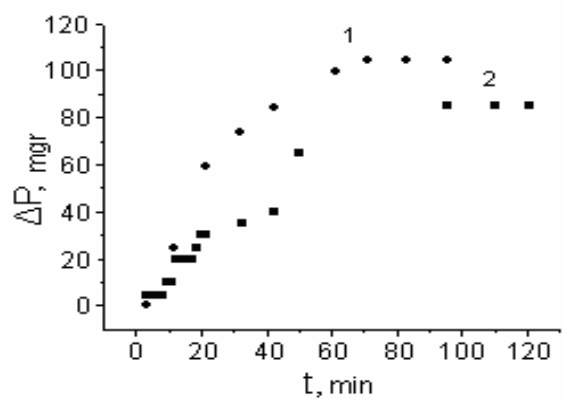

Fig.1.

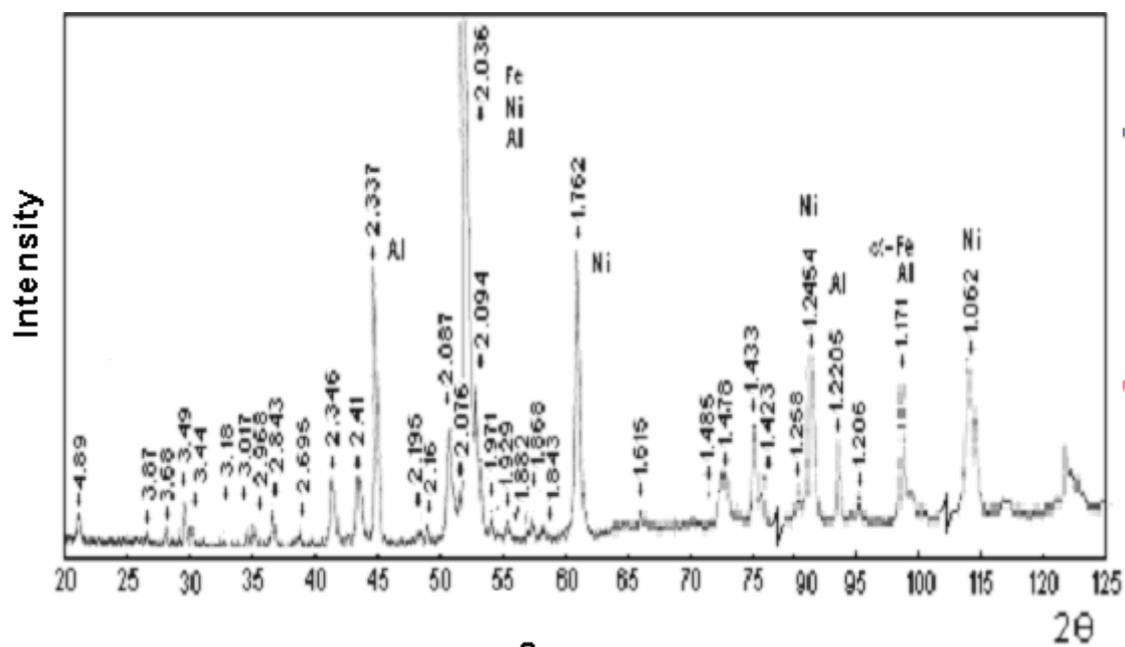

**a**

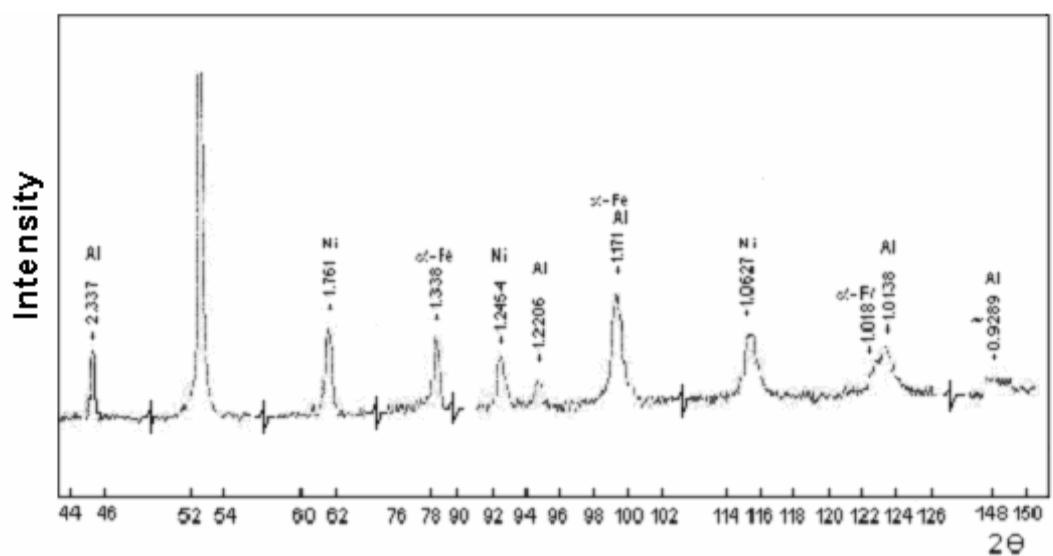

**b**

Fig.2

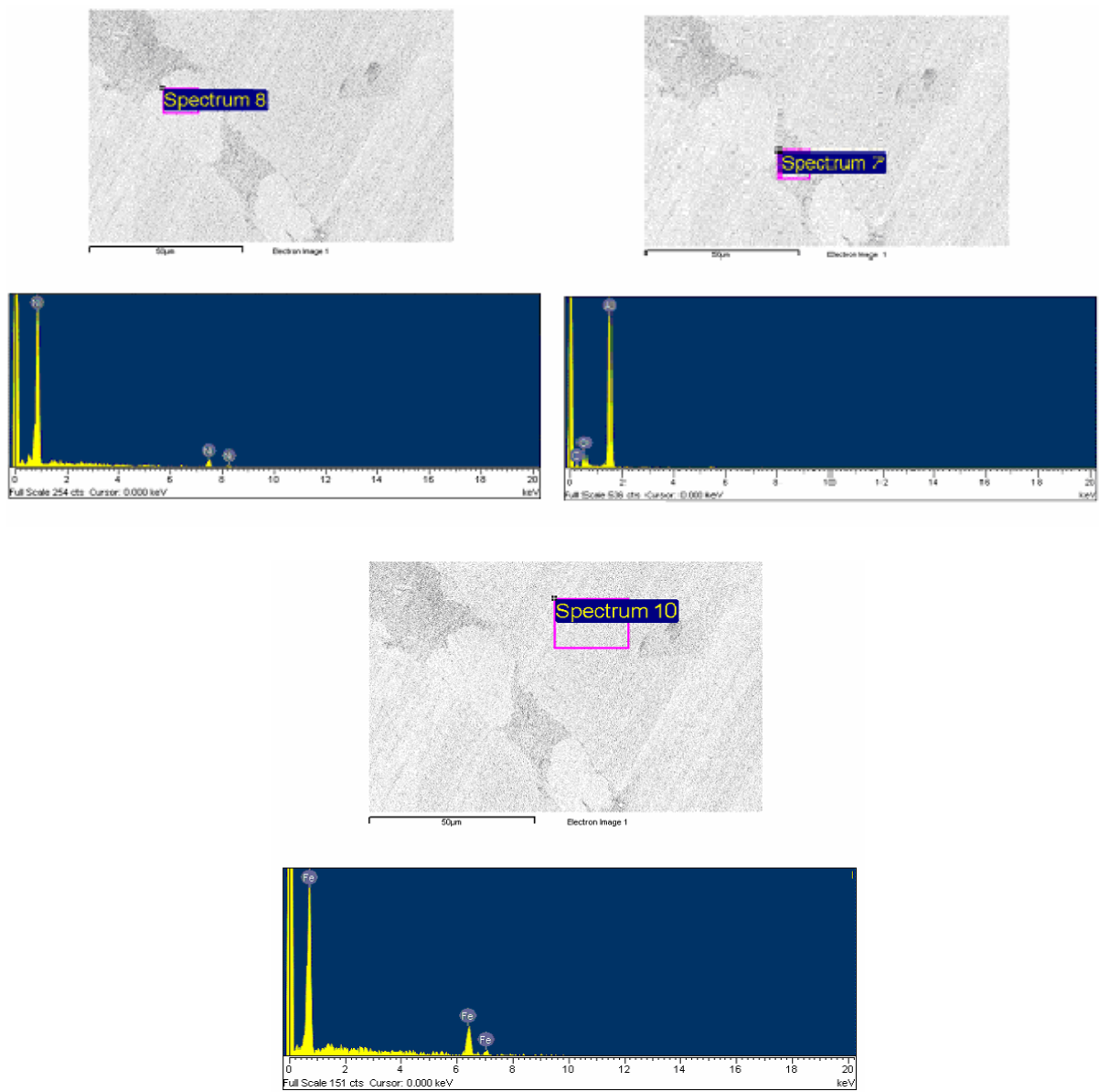

Fig. 3

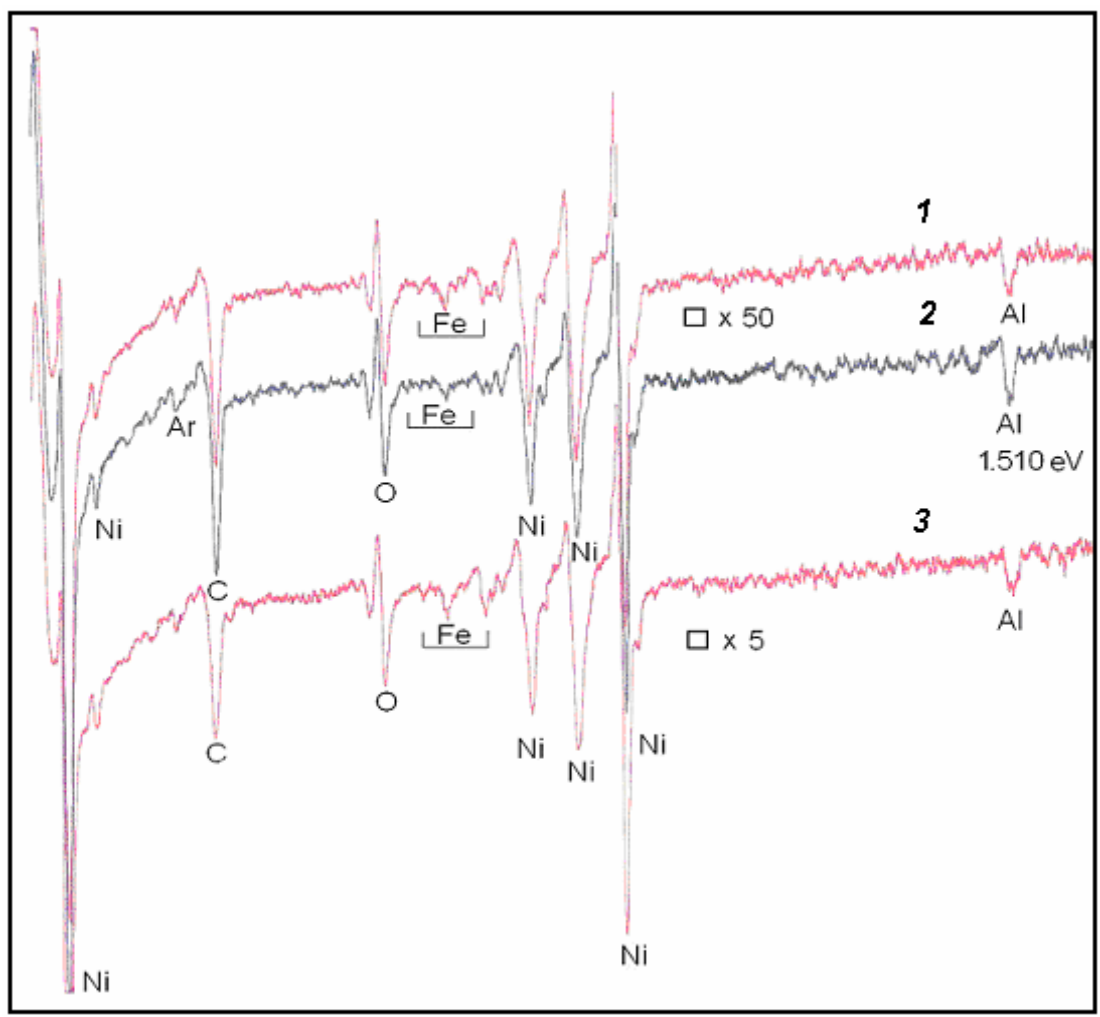

Fig. 4

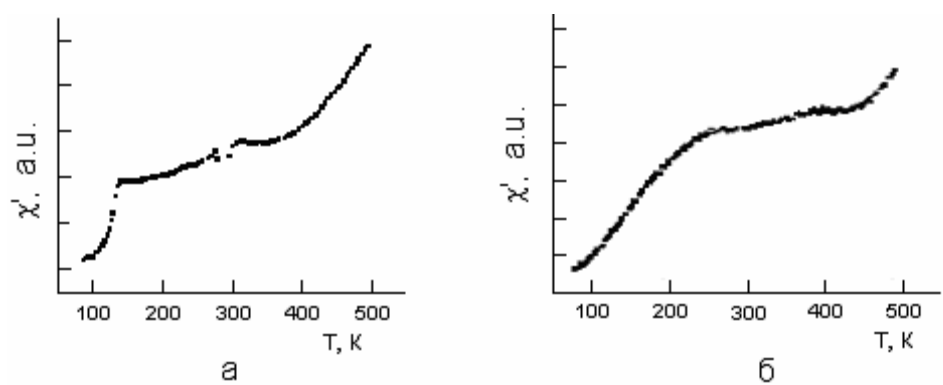

Fig. 5

FIGURE CAPTIONS

Fig. 1. Weight change dependence ($\Delta$P) of samples in alcohol on time (t) of observation. Curves (1) and (2) were obtained for a pressed sample and a sample fabricated by shook-wave compaction, correspondingly;

Fig. 2 X-ray diffraction spectra of following samples: pressed (2a) and fabricated by SWC (2b);

Fig. 3. Microstructure and spectra of the sample fabricated by the SWC method. It was used the SEM data.

Fig. 4. Auger spectra of the surface of the investigated composite Fe –Ni – Al material. 1,2,3 correspond to the different scanning areas. Curve 2 corresponds to the local value.

Fig. 5 Temperature spectra of the real part of dynamic complex magnetic susceptibility $\chi'(T)$ pressed (a) and the SWC fabricated sample (b).